\documentclass[sigconf,nonacm]{acmart}
\usepackage{graphicx}
\usepackage{float}
\usepackage{stfloats}
\usepackage{afterpage}

\AtBeginDocument{%
  \providecommand\BibTeX{{%
    \normalfont B\kern-0.5em{\scshape i\kern-0.25em b}\kern-0.8em\TeX}}}

\setcopyright{none}
\copyrightyear{2024}
\acmYear{2024}
\acmDOI{XXXXXXX.XXXXXXX}

\acmConference[ACM DocEng '24]{Make sure to enter the correct
  conference title from your rights confirmation email}{August 20--23,
  2024}{San Jose, CA}
\acmISBN{978-1-4503-XXXX-X/18/06}

\begin{document}

\title[CatalogBank: A Structured and Interoperable Catalog Dataset with a Semi-Automatic Annotation Tool (DocumentLabeler)]{CatalogBank: A Structured and Interoperable Catalog Dataset with a Semi-Automatic Annotation Tool (DocumentLabeler) for Engineering System Design}

\author{Hasan Sinan Bank}
\authornote{Corresponding author}
\email{sinan.bank@colostate.edu}
\orcid{1234-5678-9012}
\affiliation{%
  \institution{Colorado State University}
  \city{Fort Collins}
  \state{Colorado}
  \country{USA}
  \postcode{80523}
}

\author{Daniel R. Herber}
\email{daniel.herber@colostate.edu}
\affiliation{%
  \institution{Colorado State University}
  \city{Fort Collins}
  \state{Colorado}
  \country{USA}
  \postcode{80523}}


\begin{abstract}
In the realm of document engineering and Natural Language Processing 
(NLP), the integration of digitally born catalogs into product 
design processes presents a novel avenue for enhancing information 
extraction and interoperability. This paper introduces CatalogBank, 
a dataset developed to bridge the gap between textual descriptions 
and other data modalities related to engineering design catalogs. We 
utilized existing information extraction methodologies to 
extract product information from PDF-based catalogs to use in 
downstream tasks to generate a baseline metric. Our approach not 
only supports the potential automation of design workflows but also 
overcomes the limitations of manual data entry and non-standard 
metadata structures that have historically impeded the seamless 
integration of textual and other data modalities. Through
the use of DocumentLabeler, an open-source annotation tool adapted 
for our dataset, we demonstrated the potential of CatalogBank in 
supporting diverse document-based tasks such as layout analysis and 
knowledge extraction. Our findings suggest that CatalogBank can 
contribute to document engineering and NLP by providing a 
robust dataset for training models capable of understanding and 
processing complex document formats with relatively less effort 
using the semi-automated annotation tool DocumentLabeler.
\end{abstract}





\maketitle
\section{Introduction}
In the last decade, artificial intelligence (AI) has seen remarkable growth, 
starting from significant milestones such as the introduction of AlexNet in
2012 for image processing challenges \cite{krizhevsky2012imagenet} and the advent of transformers in 2017 
to address text and natural language processing tasks \cite{vaswani2017attention}. These pivotal moments,
while propelling AI forward, have also unveiled an array of technical debts and 
challenges, notably in data collection and preparation \cite{sculley2015hidden}. This pivotal observation
underscores a fundamental aspect of AI's evolution: the intricate balance between 
the innovation's requirements and the complexities it harbors.\newline
The success of AI and machine learning initiatives is deeply rooted in their 
ability to harness vast datasets, necessitating considerable investment in data 
labeling across diverse modalities, such as text, images, and others, tailored to
the demands of specific downstream tasks. This requirement accentuates an inherent
challenge: as AI solutions evolve to address more complex and varied tasks, the 
intricacies of managing and integrating these diverse data modalities escalate. 
The leap from addressing technical debts to mastering the nuances of multimodal
data underscores the need for advanced methodologies and tools that are capable 
of navigating this multifaceted landscape effectively.\newline
Despite longstanding theories that language is a fundamental aspect of consciousness 
\cite{searle2002consciousness} and discussions on its limits of perception 
\cite{wittgenstein1922tractatus}, applications of natural language 
in Computer Science before the era of GPTs were less sophisticated \cite{radford2019language}. Researchers 
in academia and industry have recognized the impact and capabilities of these 
technologies \cite{dell2023navigating, eloundou2023gpts}, and they have started to integrate this technique across a broad 
spectrum of problems, enhancing their approach to incorporating various data 
modalities.\newline
\begin{figure*}[!t]  
  \centering
  \includegraphics[width=\textwidth]{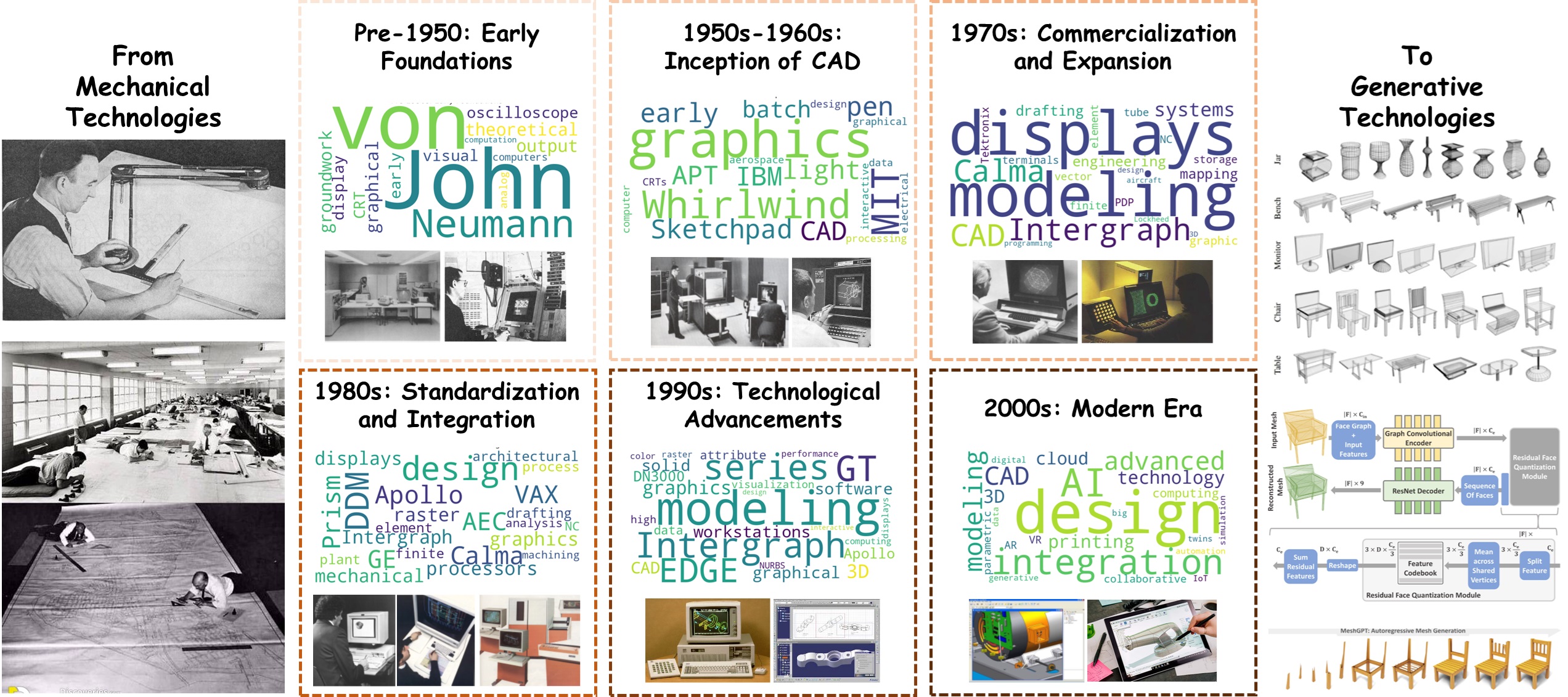}  
  \caption{Evolution of engineering design from mechanical to generative tools \cite{weisberg2008engineering, nash2020polygen, siddiqui2024meshgpt}.}
  \label{fig:0}
\end{figure*}
Naturally, from that perspective, the integration of NLP techniques
—especially transformers—with document engineering has emerged as a pivotal field for advancing
the capabilities of automated document analysis. In this regard, particularly in the
realm of design engineering, there is great promise because of the potential metadata
that engineering documents possess to interconnect the written specifications for 
function, behavior, and structure \cite{gero1990design} to the design of the system and the realization 
based on its physical features (e.g., geometry and material). As a starting point, this
paper introduces DocumentLabeler, a semiautomatic multi-modal data annotation tool 
specifically crafted to bridge the gap between textual data and product information 
within engineering documents. To leverage digitally born catalogs in the Portable 
Document Format (PDF), we take advantage of open-source tools, models, and AI frameworks
to extract product information. This approach not only enhances traditional data 
labeling but also addresses the critical limitations associated with manually entered
repositories and the interoperability of metadata for further design software integration.
We categorize our contributions in twofold. First, we present part of the CatalogBank 
dataset, emphasizing its creation from digitally born catalogs. Second, we introduce 
an open-source tool for multi-modal labeling and discuss the broader implications of the 
dataset for the document engineering domain using open-source libraries and frameworks 
(e.g., PyTorch) as the application of baseline models for efficient information extraction. 
We envision that this tool will not only support various downstream tasks, contributing 
to the advancement of document engineering, but also serve as a comprehensive resource for 
researchers and practitioners alike. This endeavor aims to foster the development of more
sophisticated and efficient tools for document analysis, thereby enriching the design engineering 
domain and beyond.
\subsection{Background}
The design process in engineering has evolved significantly over the decades. Initially, 
engineers relied predominantly on mechanical tools, progressing to electro-mechanical 
instruments, and eventually embracing digital tools with the rise of computers and 
specialized software, as shown in Fig. \ref{fig:0} \cite{weisberg2008engineering}. Each technological leap has enabled more precise and efficient 
design workflows while requiring certain cognitive effort. However,
the state-of-the-art approaches aim to offer a way to reduce this cognitive workload for the design of complex systems, 
potentially allowing engineers to explore broader and deeper aspects of design space while enhancing 
the efficiency of the process \cite{clay2023thinking}. \newline
As the engineering design process evolves, future advancements will increasingly rely on computational power to generate solutions. 
This progress depends heavily on a comprehensive understanding of interdisciplinary concepts and the ability to synthesize them 
into practical applications\cite{picard2023concept}. As a result, the impact of these advanced computational techniques on 
engineering highlights the need for comprehensive multimodal datasets to support advanced data-driven computational methods 
\cite{raina2022design, regenwetter2022deep}. \newline  
Despite progress in developing geometry-based datasets \cite{koch2019abc,willis2021fusion}, the lack of large, structured, and
multimodal datasets continues to hinder the generalizability and performance of deep learning models in engineering design. Textual 
information, in particular, is one of the essential facets for bridging this gap. As noted in various 
studies, textual data extracted from large corpus (e.g., Wikipedia) provides the
semantic context needed for effective knowledge representation and automation. 
For example, Cheong et al. \cite{cheong2016automated} implemented natural language processing techniques, 
including syntactic parsing, lexical knowledge bases, and extraction rules, to automatically extract system 
structure knowledge (e.g., objects' function) from text and compared their results against repositories with 
manually entered information, such as \cite{szykman1999nist} and \cite{bohm2006introduction}. 
Cheong et. al. highlighted that the repositories with manually entered information have a limitation 
in terms of scalability. Other research studies on natural language processing, such as WordNet \cite{miller1990introduction}, 
ConceptNet \cite{speer2017conceptnet}, BLine \cite{shi2017data}, and TechNet \cite{sarica2020technet}, 
focus on successfully forming a semantic network and design representation without 
incorporating the spatial information like 2D or 3D geometry for design, planning, or 
manufacturing purposes \cite{sarica2023design}.\newline
From this point, one of the ideas to consider is the utilization of knowledge extraction 
methods to combine the spatial information (image, 3D geometry, 2D technical drawing, etc.) of an object 
and relevant information related to this object from external references. However, there is
a lack of ground truth for generating knowledge from a scalable external reference and 
injecting extracted knowledge to the point where it makes sense to use
with spatial information for its designed purpose. \\ 
Similar to \cite{cheong2016automated}---in terms of targeting textual information---Williams \cite{williams2020comparing}  
aims to collect the attribute- and form-based information with the associated geometry data from the web. However, their approach is 
specific to the vendor's page layout and is not generalizable enough to translate to other vendors. 
Although there are a plethora of libraries and web platforms, such as Beautiful Soup, Selenium,
and browse.ai, for web scraping, the main issue is the complexity of dynamic web pages (e.g., 
JavaScript-based) or using third-party services that website owners employ to prevent data mining. Even workarounds such as those
using sitemap files (e.g., XML, .xfm, etc.) from robot.txt of the web pages, utilization of residential
IP proxies or UI-based automation tools (e.g., Autokey, Autohotkey, Autoit, etc.) still complicate
the potential for generalizing the method to extract information from the web. \newline
As we can see from the literature, significant efforts have been made to improve the modality, quality, 
and quantity of engineering design datasets, much work remains to be done to address the limitations of existing 
ones. The integration of new tools, such as DocumentLabeler, with the standards (e.g., ISO 10303) will be beneficial 
for advancing the use of other modalities with the geometry in engineering system design and new approaches to knowledge 
extraction and natural language processing will be essential for realizing the full potential of these tools. 
\section{The Description of the CatalogBank Dataset and DocumentLabeler} 
In this section, we elaborate on the details of the presented dataset 
architecture and provide more information regarding the document 
part of the dataset and supplementary information of this paper. We 
outline the essential components of the presented dataset for 
enabling multimodality, including the appropriate size, accurate baseline data, 
variable data structure, and scalability. An adequate number of data 
is required for statistical significance, whereas inaccurate data 
can result in an improperly trained model. The sampling of data and 
translation must not lead to additional errors, and the dataset 
should be well-documented with the necessary scripts to facilitate 
data filtering and wrangling. Furthermore, the dataset should 
contain different types of data from various categories to ensure 
data heterogeneity, and each type should have sufficient existing 
data distribution to prevent bias in any trained model.
\subsection{Document Dataset: CatalogBank}
When we look into the standard design workflow of a design engineer, 
we see the utilization of catalog-like web pages or tools by design 
engineers to extract product information and geometric models from 
conceptual design to the end of the design process. Therefore, we 
developed the idea of utilizing digitally born catalogs in native 
Portable Document Format (PDF) \cite{gabdulkhakova2012document} to extract product information using 
NLP techniques. Here, we consider the digitally born document as 
created in a word processor or vector-based design software (e.g., 
CorelDraw, Inkscape, Adobe Illustrator, etc.) and stored as a PDF 
consisting of the document’s information as metadata (e.g., by image 
and character, etc.) that does not need an additional process for 
Optical Character Recognition (OCR) to extract the characters or 
image of the document. In document engineering, there have been new 
datasets as well as the utilization of new techniques for layout 
analysis and knowledge extraction, such as the analysis of document 
layout (e.g., LayoutLM \cite{xu2020layoutlm}), information 
extraction from tables (e.g., TableBank \cite{li2020tablebank}, 
DeepDeSRT \cite{schreiber2017deepdesrt}, TaPas 
\cite{herzig2020tapas}], Rethinking Table 
\cite{qasim2019rethinking}), and information extraction from 
documents (e.g., Donut \cite{kim2022ocr})---or material safety data sheets (MSDS)
\cite{fenton2021engineering}.\newline
\begin{figure}[!t]
  \centering
  \includegraphics[width=0.475\textwidth]{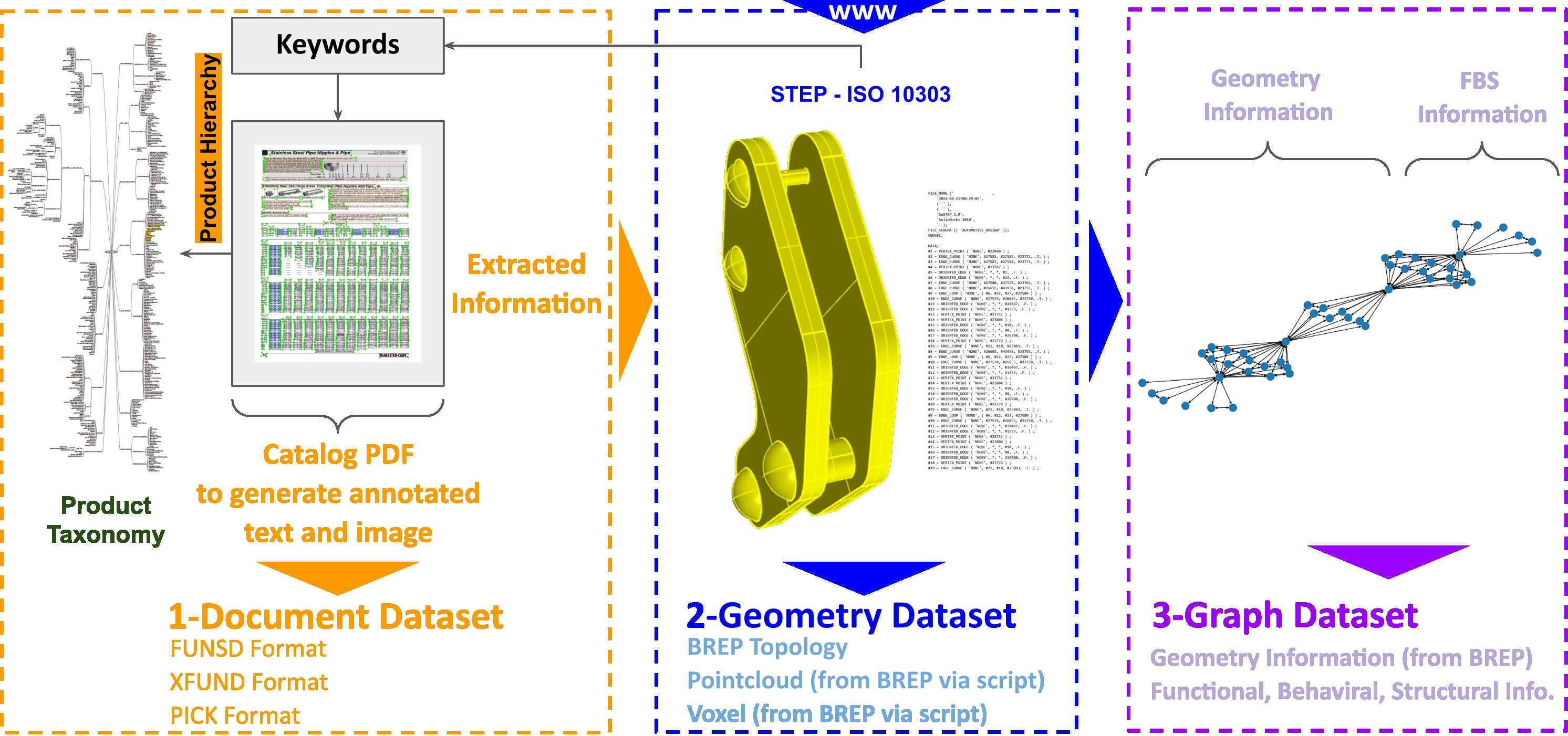}
  \caption{A sample for presenting the overall view of the complete CatalogBank dataset from McMaster Carr v125.}
  \Description{A woman and a girl in white dresses sit in an open car.}
  \label{fig:2}
\end{figure}
The techniques presented in these research studies heavily rely on 
the datasets’ domain –or the similarities of the layout between the 
dataset and the target applications of the documents. Some of these 
well-known datasets or corpora include Wikipedia 
\cite{denoyer2006wikipedia}, PubLayNet\cite{zhong2019publaynet}, 
FUNSD\cite{jaume2019funsd}, XFUND\cite{xu2021layoutxlm}, 
DocBank\cite{li2020docbank}, and SROIE \cite{huang2019icdar2019}, 
among others. \newline
The deep learning models that needed to be trained 
standardized their data input based on some of these dataset’s 
format. In the case of the dissimilarity between the trained and 
target document, these aforementioned models needed to be fine-tuned 
or re-trained with documents similar to the target document. For 
example, in a lot of cases the people who are developing the models 
are using databases from public journals. That approach gives the 
ability for the model to learn similar layouts and the information 
within it. However, catalogs are not exactly the same as these 
documents. Therefore, a new dataset based on the information from 
the engineering catalogs would be beneficial. \newline
Given the limited research of knowledge extraction to combine the 
design geometries to their specifications and the progress made in 
the NLP domain (especially in document engineering), we propose a 
new dataset called CatalogBank that state-of-the-art algorithms can utilize
to combine NLP and other advanced geometry
algorithms. By using digitally born versions, we were able to 
generate an image and annotate every minutiae of the data existing 
from a native PDF catalog to overlay the information on the image 
for multimodality in further processes. We can also test our dataset 
with different baseline algorithms to appreciate the usefulness 
of the dataset for different document-based downstream tasks. We 
have included images in our dataset to enable testing of document 
engineering methods in a way and presented a sample with the 
architecture from McMaster Carr v125 in Fig. \ref{fig:2}. With the 
provided software solution DocumentLabeler and the scripts in this 
paper’s GitHub repositories \cite{bankh_CatalogueBank}, 
\cite{bankh_DocumentLabeler}, and work-in-progress ones, we created thousands of parts with 
their functional and physical properties incorporated into a graph 
structure and taxonomy with hundreds of categories with a total 
of millions of features from various vendors for domains to cover the whole CatalogBank dataset (Fig.\ref{fig:2}). While we are continuously growing the content of CatalogBank, 
as presented for this paper, one can find a total of 11,984 pages from the catalogs (Misumi, 
Newark, Thorlabs, McMaster-Carr, 8020, and Grainger as shown in Fig.
\ref{fig:4}). By using the information in a standard catalog, we 
ensured that the generated data is not relying on non-standardized or informal 
expertise (e.g., "Wisdom of the Crowds" \cite{burnap2015crowdsourcing}) or opinions, the information rather relies on standard 
engineering data. We provide the details of one of the catalogs as a 
brief summary in Table \ref{tab:my_label}.
\begin{table}[!h]
  \centering
  \caption{The partial content of the dataset CatalogBank from 
  Thorlabs v21}
  \label{tab:my_label}
   \begin{tabular}{p{1.65cm} p{1.3cm} p{1.8cm} p{0.95cm} p{0.75cm}}
    \toprule
    \textbf{Vendor} & \textbf{Document} & \textbf{Products/CAD} & \textbf{Images} & \textbf{Graph} \\
    \midrule
    Thorlabs v21 & 1,803 pgs & 29,329/24,096 & 361,440 & 24,096 \\
    \bottomrule
  \end{tabular}
\end{table}
\begin{figure}[!b]
  \centering
  \includegraphics[width=0.475\textwidth]{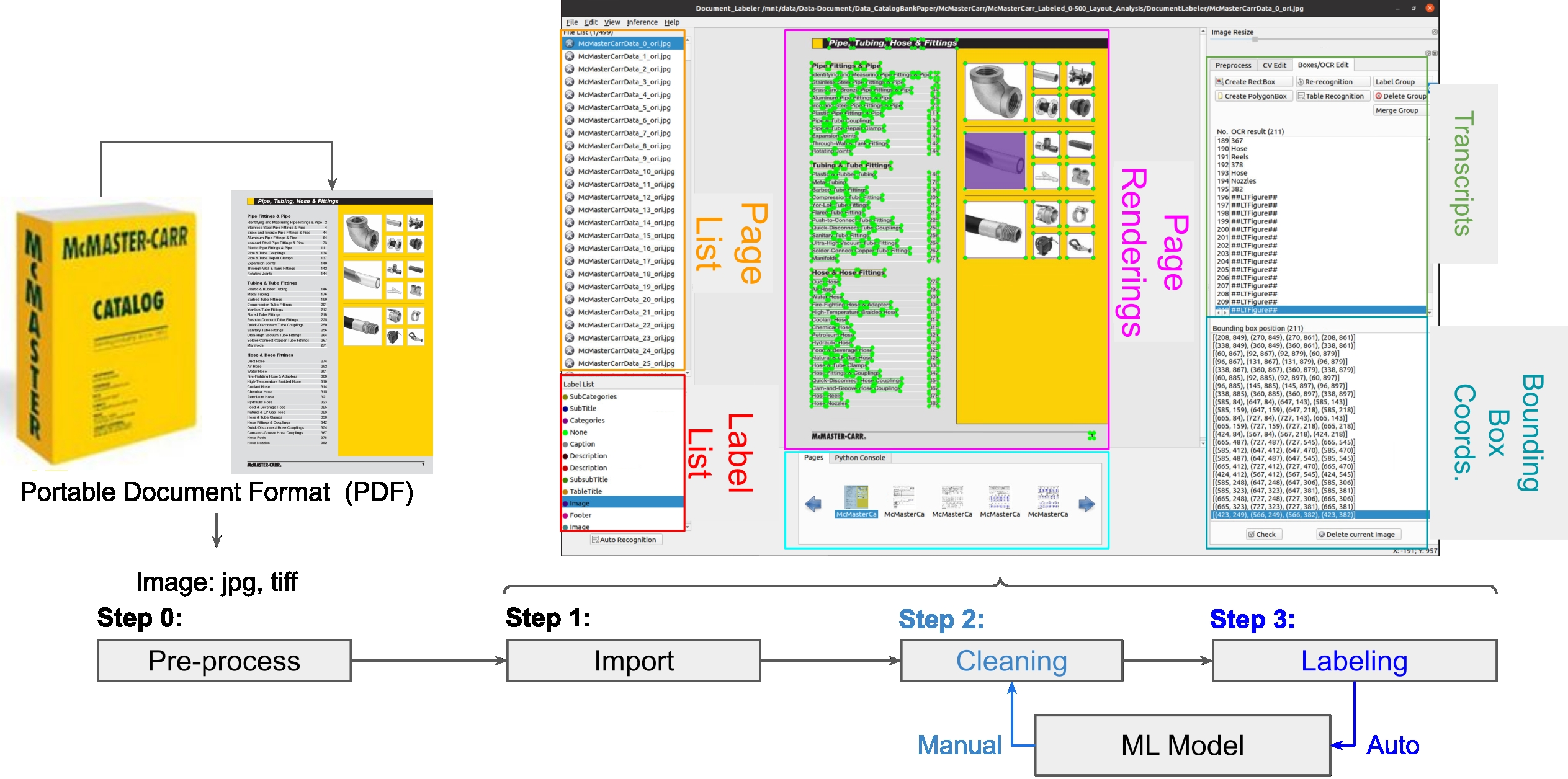}
  \caption{The details of document annotation and information extraction workflow Document Dataset (ML Model: PICK \cite{yu2021pick} or others).}
  \label{fig:3}
\end{figure}
\subsection{Annotation Tool: DocumentLabeler}
\begin{figure}[!h]
  \centering
  \includegraphics[width=0.475\textwidth]{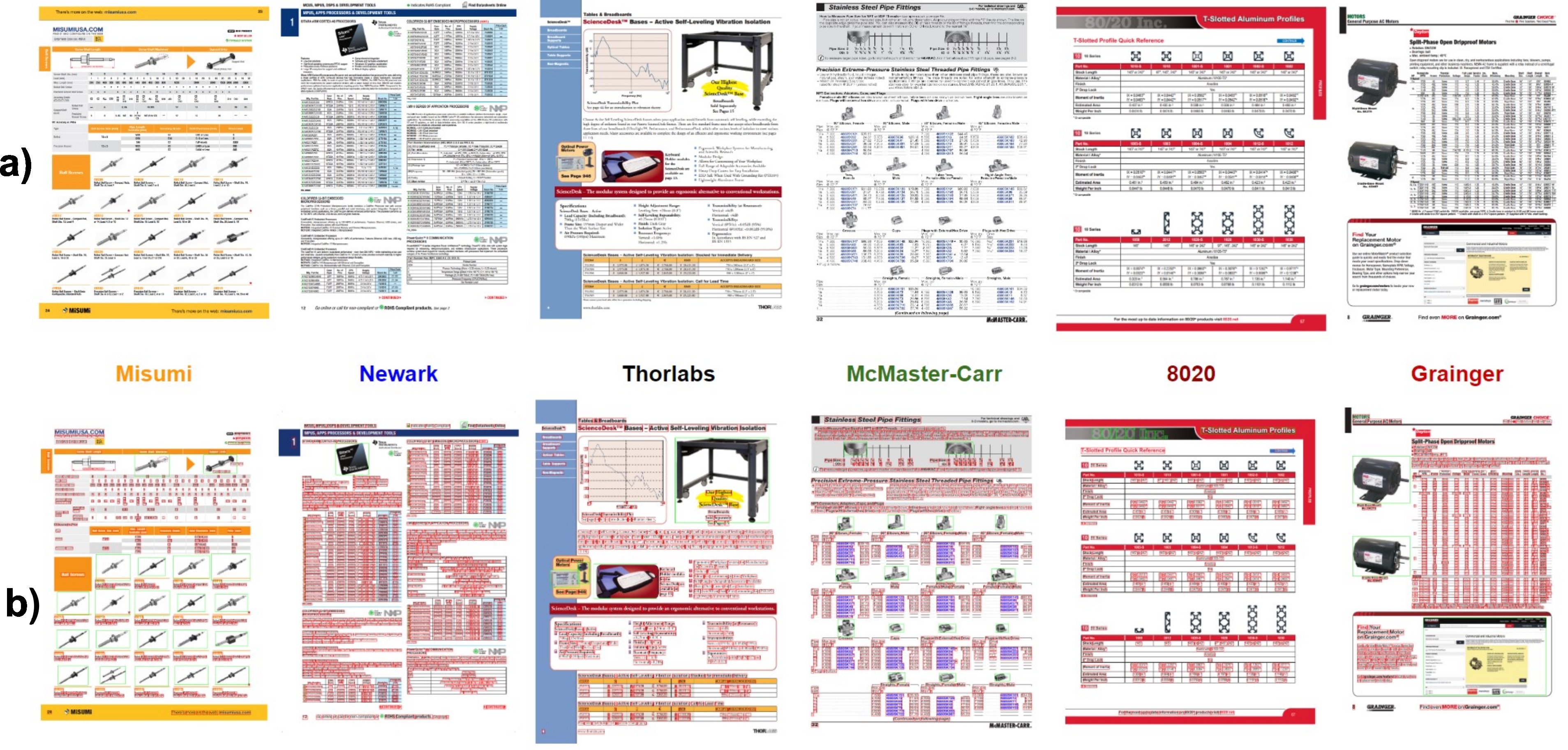}
  \caption{\textbf{a)} Digitally-born Catalogs in PDF and \textbf{b)} after preprocessing (Peruse of Step 0 from Fig. \ref{fig:3}) from well-known vendors such as Misumi, Newark, Thorlabs, McMaster-Carr, 8020, and Grainger, respectively.}
  \label{fig:4}
\end{figure}

\begin{figure*}[!t]
  \centering
  \includegraphics[width=\textwidth]{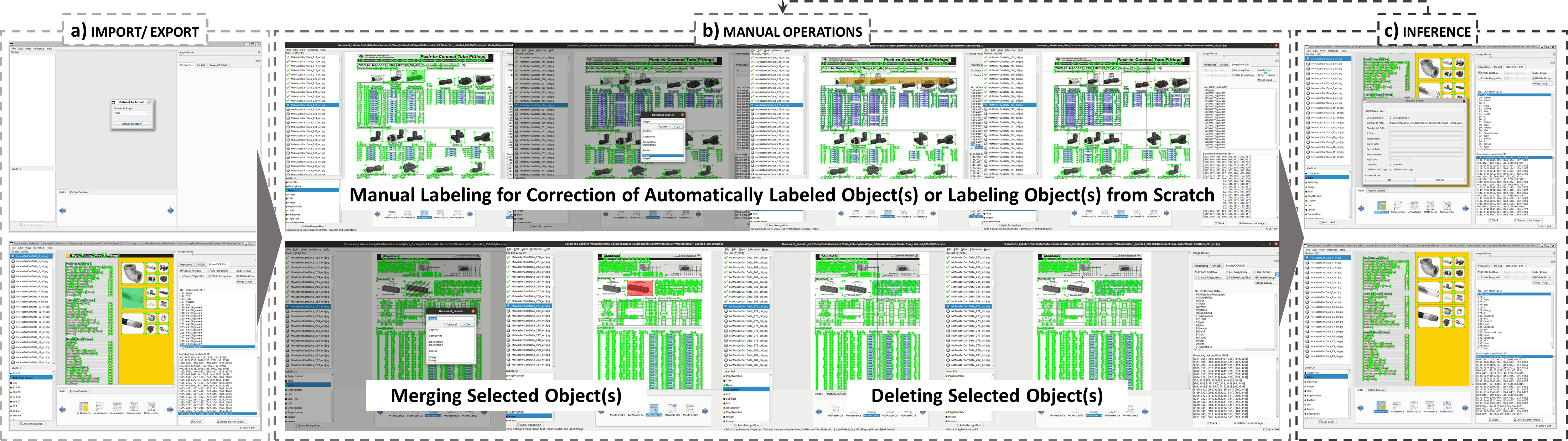}
  \caption{\textbf{a)} Importing Data, \textbf{b)} Manual Operations (Labeling, Merging, or Deleting), and 
  \textbf{c)} Inference on Selected Model}
  \label{fig:5}
\end{figure*}

Since 2015, ACM DocEng has published approximately 282 publications, comprising 176 short papers 
and 106 research articles, that illuminate the forefront of challenges and solutions in document 
engineering. Among these, as per ACM's search results, 118 publications delve into artificial intelligence,
covering both machine learning and deep learning methodologies. It is observed that many authors have 
harnessed their unique tools for similar analytical tasks as those explored by peers, integrating their 
findings directly into a specific software solution while using well-known libraries or generic frameworks. 
Transitioning from these broader contributions, our own investigation
specifically into the realm of data labeling for design engineering problems uncovered a notable gap: the lack
of a multimodal data labeling tool that is both open-source and features an open architecture conducive to 
adaptation across various frameworks. This gap points to an essential requirement—a platform that not only 
supports an open architecture, allowing users to author their own code or import an existing one, but is 
fundamentally open-source. Despite various attempts to address the needs of either single or multiple data 
modalities, this critical criterion of openness remains unfulfilled. Bridging this gap, we identified existing
data annotation tools, such as Doccano \cite{nakayama2018doccano}, Prodigy, Supervisely, SageMaker Ground Truth, and UBIAI, which 
encounter difficulties in providing a free, offline, open-source, and open-architectural solution for multimodal
documents, such as those containing both images and text to combine with other data modalities. Recognizing the 
demand for a software tool that is not only free of charge but also community-driven and privacy-focused, we 
leveraged PPOCRLabel (based on LabelMe \cite{russell2008labelme} and LabelImg \cite{Tzutalin2015}) as a foundation. 
We have enhanced its functionality to improve user experience through features like multi-text manipulation, 
including a Python console to interact with the tool programmatically during run-time, multi-object deletion, 
labeling, and merging. We have updated the language of the user interface. Additionally, we have improved 
its connectivity with other tools, ensuring better compatibility with standard NLP libraries and interfaces 
to different machine learning frameworks, such as PyTorch, as opposed to PaddleOCR based on Baidu's PaddlePaddle. 
This enhanced tool is introduced under a new name: DocumentLabeler, as shown in Fig. \ref{fig:3}. \newline
We illustrate a process flow for transforming a PDF catalog into a 
labeled dataset for machine learning: converting the catalog into 
images, importing these images into a data system, cleaning up the 
data through both manual and automated means, and finally, annotating 
the data with labels for training an ML model. This systematic 
approach is designed to ensure that the dataset is accurate and 
structured for effective machine learning utilization. \newline
\textit{Pre-process Documents:} To ensure generality, we converted a digitally-born document to an image to form the ground truth 
dataset, as shown in Fig \ref{fig:4}. \newline
One of the potential issues with generated image-based catalogs is 
the design complexities of documents, such as color contrast, the 
distance between shapes and text, and other design factors during 
OCR. To avoid this issue, it is generally recommended to use a 
simple preprocessing workflow that 
involves resizing, binarization, and Otsu's thresholding for these 
generated images whenever necessary. These steps will clear the 
background based on the histogram of the image and result in the 
best value of the threshold to separate dark and 
light regions of the image. Furthermore, there are research studies 
\cite{lins2021binarisation} and competitions \cite{lins2023quality} 
that aim to find the best combination of preprocessing operations 
with traditional algorithms and neural network-based methods. 
effective in the case of complicated catalog backgrounds. 
However, in our dataset, we did not need to do these additional
steps other than resizing and text filtering via regex due to the 
existing character and image of the digitally born catalogs’ data 
and relatively limited complexity of the pages, as shown in the Fig. 
\ref{fig:4} and shared pre-process scripts on GitHub repository \cite{bankh_CatalogueBank}.\newline
\textit{Import/ Open or Export Documents:} There are many dataset formats 
that have become de-facto standards in document analysis and engineering. 
Therefore, in DocumentLabeler, we have included four different dataset 
formats for importing data: PICK, DocBank, XFUND, and FUNSD. Consequently,
researchers accustomed to these dataset structures can directly import their
documents into DocumentLabeler for further processing via File > Import and
similarly, the user can export their work to the target format via 
File > Export, as shown in Fig \ref{fig:5}a.\newline
\textit{Clean Imported or Opened Documents:} In a lot of cases, the data 
that is imported is not yet labeled, or the bounding boxes that represent 
the token groups might not be correctly identified. Therefore, a manual 
cleaning process might be necessary to merge the character-based bounding 
boxes to form the word-based ones or grouping the bounding box objects
from word to sentence or paragraphs or captions and images into a single object, 
as shown in Fig. \ref{fig:5}b. This step would be necessary in the case of 
potential errors or requirements in the pre-process scripts or the target 
machine learning model. \newline
\textit{Labeling of the Documents:} After proper cleaning of the 
document, one can run existing labeling models integrated into the 
software or manually label the documents. Our UI and short-cut 
enhancements with the manual labeling step shorten the manual 
labeling cycle from 30 minutes to a few minutes per page without any 
automation. In the event of automatic labeling, we can always 
utilize the tools that are developed for manual cleaning and 
labeling for the correction of errors during the auto-labeling 
process, as shown in Fig. \ref{fig:5}c.\newline
The use of the DocumentLabeler does not require an internet 
connection. Therefore, the labeling can be accomplished on the 
premises while ensuring data privacy and security. 
\section{Some Experiments with DocumentLabeler and CatalogBank}
We utilized a baseline algorithm to showcase the versatility of the 
CatalogBank’s document dataset and DocumentLabeler. For document-related
tasks as part of layout analysis and information extraction, 
we implemented PICK (Processing Key Information Extraction from 
Documents using Improved Graph Learning-Convolutional Networks)
\cite{yu2021pick} as a baseline model on both the DocBank and 
CatalogBank datasets. Each dataset was trained for a full page and 
selected number of tokens.  \newline
We adopted the PICK framework as the baseline of our experimental setup. 
Our decision was motivated by  PICK's demonstrated proficiency in extracting information from complex document layouts through a synergistic combination of transformers, graph learning, and convolution operations \cite{huang2019icdar2019}. 
\newline
\begin{figure}[!b]
  \centering
  \includegraphics[width=0.475\textwidth]{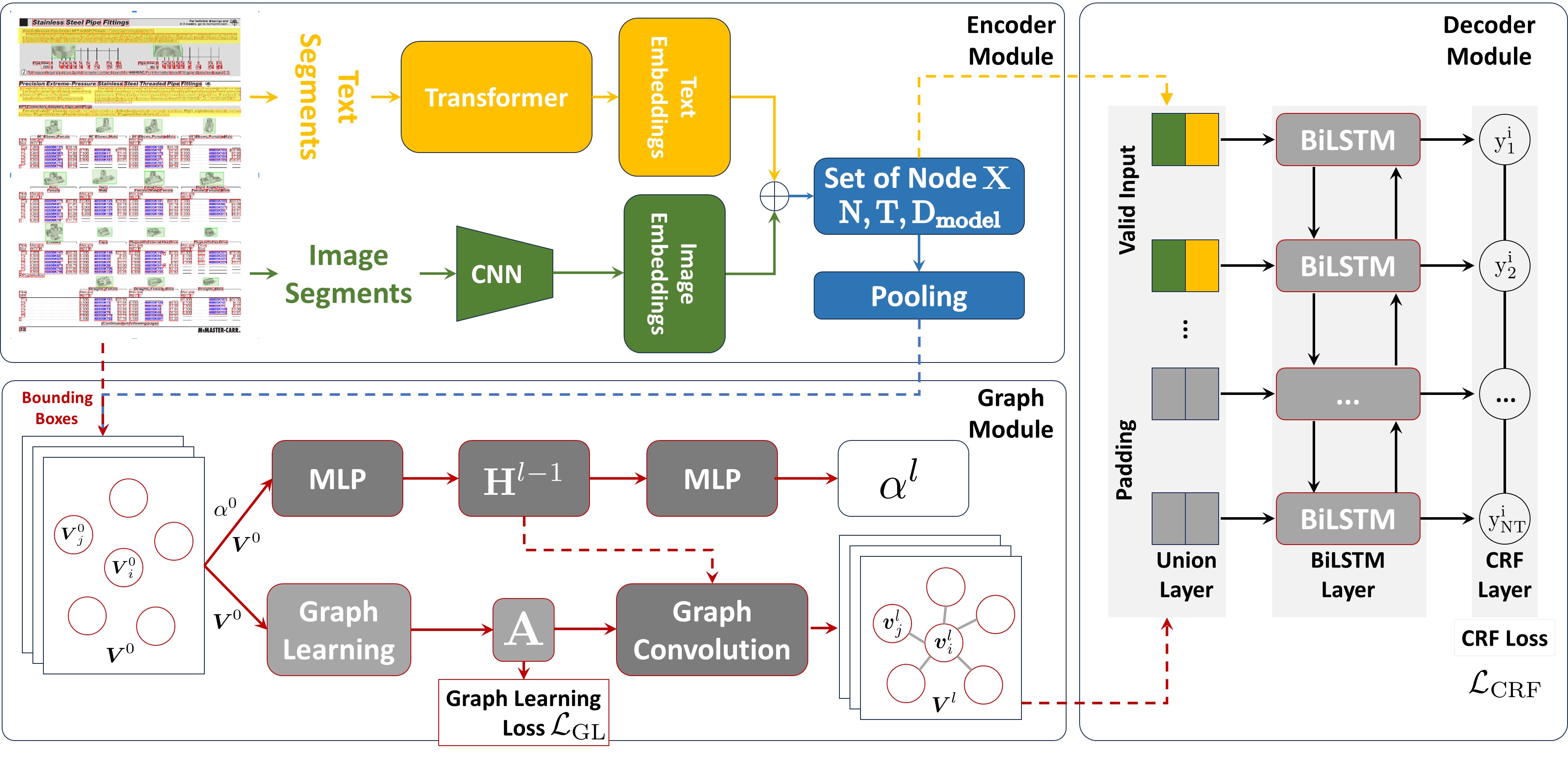}
  \caption{The architecture of PICK \cite{yu2021pick}}
  \label{fig:6}
\end{figure}
In Fig.~\ref{fig:6}, we provide the architecture of the method,
which incorporates an encoder, graph module, and decoder. The PICK 
framework's architecture is characterized by node embeddings with in the 
$l$-th graph convolution layer, where $\mathbf{\alpha}^l$ signifies the embeddings 
of relationships and $\mathbf{H}^l$ delineates the concealed characteristics 
shared amongst the nodes \( v_i \) and \(v_j \) in that particular layer. The 
matrix $\mathbf{A}$ serves as a pliable adjacency matrix. The symbols $\mathrm{N}$, $\mathrm{T}$,
and $\mathrm{D_{model}}$ represent the count of sentence segments, the 
upper limit of sentence length, and the scale of the model's dimensions, respectively. 
Additionally, \( \oplus \) is used to indicate an element-wise summation. \newline
The models are trained from scratch using Adam \cite{kingma2014adam} as the optimizer to simultaneously minimize CRF 
(Conditional Random Field) and graph learning losses, and the batch size is 12 during the training of the model. The 
learning rate is set to $10^{-4}$ over the whole training, with a step decay by a factor of 0.1 every 30 epochs. We 
use dropout with a ratio of 0.1 on both BiLSTM and the Transformer. The model is trained for 100 epochs
(CatalogBank) with approximately 3 minutes per epoch and 60 epochs (DocBank) with approximately 245 minutes per epoch. 
Early stopping is employed with patience of 20 epochs for CatalogBank and 5 epochs for DocBank, stopping training if 
the monitored metric does not improve within these periods. At the inference phase, the model directly predicts every 
text segment that belongs to the most probable entity type. The reader can find these details in each trained model's
log and config files \cite{bankh_DocumentLabeler}.\newline
\begin{table}[!t]
  \centering
  \caption{Performance Metrics of PICK on DocBank and CatalogBank Datasets during training for layout analysis}
  \label{tab:performance_metrics}
  \begin{tabular}{lcccc}
    \toprule
    & \textbf{mEP} & \textbf{mER} & \textbf{mEF} & \textbf{mEA} \\
    \midrule
    DocBank & 0.91 & 0.91 & 0.91 & 0.91 \\
    CatalogBank & 0.99 & 0.99 & 0.99 & 0.99 \\
    \bottomrule
  \end{tabular}
\end{table}
The initial dataset of CatalogBank (McMasterCarr v125 catalog) to train PICK is annotated manually for the first 500 
pages out of 3,378. DocumentLabeler's UI and associated modifications and shortcuts were quite effective in this regard. 
On the other hand, the DocBank dataset (approx. 500,000 pages) is already annotated as presented in \cite{li2020docbank}. 
During our training and validation, we utilized a wrapper of a custom dataset class for PyTorch (\texttt{torch.util Dataset}) with 
a 4:1 ratio (400 pages of training and 100 pages for validation). For inference (testing), the rest of the document can be 
utilized from McMaster Carr's 2,878 pages or the rest of the shuffled CatalogBank dataset to test the generality of the 
inference. The same training and validation ratio is applied for the training of the DocBank-based model (4:1 of 10,000 
pages). We also provide an alternative approach in our GitHub repository where we shuffle the overall dataset and train 
the model with the same hyperparameters, number of pages, and train-to-validation ratio. \newline
Total training for the CatalogBank data was completed in 100 epochs 
in approximately 315  minutes using a custom-designed workstation\footnote{Bank, 
H.S.,. "Notes and Tools for GPU Computation." *GitHub*, 
github.com/bankh/GPU\_Compute. Accessed 18 Apr. 2024.} for AMD Graphical Processing Units. Vectorcraft equipped with AMD 
Threadripper 3955X, 256GB of RAM, and 7 AMD Instinct GPUs totaling 112 GB of vRAM on Ubuntu 20.04 Focal in a Docker 
container with ROCm 4.0.1, Python 3.8, and PyTorch 1.8, trained the full page (roughly 
between 700 and 1000 objects per page) for 500 pages with 6 GPUs. The associated training times are depicted in 
detailed log files of training are available on the GitHub repository 
\cite{bankh_DocumentLabeler}. The same system (with 7 GPU) was 
utilized for 10,000 pages over 60 epochs, taking 14,818 minutes (apx. 10.3 days). The results are 
also provided in Table \ref{tab:performance_metrics} and a sample 
structure of a catalog page with the total number of label occurrences in Fig. \ref{fig:6}. \newline
\begin{table}[!b]
  \centering
  \caption{Detailed training performance of PICK by document element types on CatalogBank dataset with specific labels for layout analysis}
  \label{tab:detailed_performance}
  \begin{tabular}{lcccc}
    \toprule
    \textbf{Element Type} & \textbf{mEP} & \textbf{mER} & \textbf{mEF} & \textbf{mEA} \\
    \midrule
    Image & 0.921 & 0.901 & 0.911 & 0.901 \\
    SubsubCategories & 0.705 & 0.896 & 0.789 & 0.896 \\
    Categories & 0.909 & 0.909 & 0.909 & 0.909 \\
    PageNumber & 0.458 & 0.407 & 0.431 & 0.407 \\
    Description & 0.995 & 0.999 & 0.997 & 0.999 \\
    TableTitle & 0.878 & 0.915 & 0.896 & 0.915 \\
    SubCategories & 0.730 & 0.979 & 0.836 & 0.979 \\
    Table & 0.993 & 0.996 & 0.995 & 0.996 \\
    Title & 0.982 & 0.996 & 0.989 & 0.996 \\
    SubTitle & 0.983 & 0.987 & 0.985 & 0.987 \\
    List & 0.995 & 0.996 & 0.995 & 0.996 \\
    SubsubTitle & 0.971 & 0.927 & 0.949 & 0.927 \\
    \textbf{Overall} & 0.990 & 0.994 & 0.992 & 0.994 \\
    \bottomrule
  \end{tabular}
\end{table}
Similar to \cite{guo2019eaten}---and consistent with \cite{yu2021pick}---, to evaluate the efficacy 
of our experiments with the PICK framework on the CatalogBank dataset, we employed several key metrics: 
Mean Entity Precision (mEP), Mean Entity Recall (mER), Mean Entity F-1 Score (mEF), and Mean Entity Accuracy (mEA). 
These metrics offer a detailed perspective on the model's performance across various dimensions of information extraction tasks. In the equations
that follow, $y^i$ represents the predicted text, and $g^i$ represents the target text of the i-th entity. 
$I$ is the number of entities, and $\mathbb{I}$ is used to denote the indicator function that returns 1 if $y^i$ 
is equal to $g^i$, and 0 otherwise.\newline
\textit{Mean Entity Precision (mEP):} This metric quantifies the accuracy of the extracted entities by calculating
the ratio of correctly extracted entities to the total entities extracted by the model.
\begin{equation}
m E P=\sum_{i=0}^{I_p-1} \mathbb{I}\left(y^i==g^i\right) / I_p
\label{eq:mep}
\end{equation}
\textit{Mean Entity Recall (mER):} This metric assesses the model's ability to identify and extract all relevant 
entities from the document.
\begin{equation}
m E R=\sum_{i=0}^{I_g-1} \mathbb{I}\left(y^i==g^i\right) / I_g
\label{eq:mer}
\end{equation}
\textit{Mean Entity Accuracy (mEA):} This metric evaluates the overall accuracy of the entity extraction,
considering both correctly extracted entities and those that were incorrectly extracted or missed.
\begin{equation}
m E A=\sum_{i=0}^{I-1} \mathbb{I}\left(y^i==g^i\right) / I
\label{eq:mea}
\end{equation}
$I_p$ is the number of non-null predicted entities, and $I_g$ is the number of non-null target entities. 
When both the prediction and target are null, the indicator function returns 0.\newline
\textit{Mean Entity F1 Score (mEF)}: The harmonic average of mEP and mER.\newline
The results showcased in Table \ref{tab:performance_metrics} and \ref{tab:detailed_performance} underscore
the effectiveness of the PICK framework in accurately extracting layout information across various document
elements, such as Tables, Title, SubTitle, and Images. The high scores in the mEP, mER, mEF, and mEA metrics
affirm the PICK framework's capability to handle the complexities inherent in the CatalogBank dataset. 
However, a deeper analysis would be beneficial using the shuffled version of the overall CatalogBank dataset.
Regardless of the model used in this experiment, these results highlight our approach's effectiveness in 
enhancing the extraction of design data information for automating design engineering processes and semi-automatic
data labeling through the DocumentLabeler tool.
\section{Conclusion and Future Work}
In this study, we introduced the document aspects of CatalogBank dataset and the DocumentLabeler tool. The CatalogBank dataset has been
curated to support the automation of design engineering processes, bridging the gap between textual descriptions and
other data modalities related to engineering design catalogs. Simultaneously, we have presented DocumentLabeler, a 
semi-automatic data labeling tool designed to facilitate the annotation of complex document formats. This tool 
represents a step towards simplifying the traditionally labor-intensive and time-consuming process of data labeling, 
offering a user-friendly interface that accommodates multimodal data input. \newline
A key component of our exploration involved utilizing a PyTorch-based framework (e.g., PICK) within the context of 
DocumentLabeler for layout analysis. This integration not only showcased the potential of the dataset
and the efficiency of the tool but also highlighted the potential for sophisticated models to 
advanced information extraction from complex document layouts. \newline
Moreover, we foresee implementing features such as collaborative labeling \cite{tomasi2013collaborative}, which allows multiple annotators to work
concurrently from the same internal --or local-- network (e.g., intranet), enhancing the speed and accuracy of the annotation process. We also consider to focus
on seamlessly interfacing the preprocessing scripts and computer vision aspects (for enhanced OCR) within the user interface of
DocumentLabeler, ensuring a smooth and intuitive workflow for users. Furthermore, integrating the utilization of the integrated 
Python console for accessing and utilizing as a prompt for different Large Language Models (LLMs) during runtime will significantly expand the tool's 
adaptability and functionality in data handling and processing (e.g., key information extraction) by using state-of-the-art approaches. Finally, the development of a 
web interface for our codebase brings an opportunity to make our tools and dataset more accessible to the wider research and development community, fostering 
collaboration and innovation in document engineering and NLP. In future publications, we will introduce geometry models and graph modalities of CatalogBank with 
associated functionalities of DocumentLabeler in more depth. \newline
As we conclude, the contributions of this study—ranging from the presentation of the document engineering related aspects for 
CatalogBank dataset to the introduction of DocumentLabeler, and as a PyTorch application of the PICK framework—lay a foundation for future
research in document engineering. Our future efforts to integrate collaborative labeling, enhance the user interface, and 
develop a web interface represent forward steps in making more advanced document processing tools more accessible and 
effective. By sharing our implementation on GitHub repositories \cite{bankh_CatalogueBank} and \cite{bankh_DocumentLabeler}, we hope that our work will inspire 
further developments and applications in the realm of document engineering, engineering system design, and beyond.

\bibliographystyle{ACM-Reference-Format}
\bibliography{sample-base}

\end{document}